\documentclass{article}

\usepackage{PRIMEarxiv}

\usepackage[utf8]{inputenc} 
\usepackage[T1]{fontenc}    
\usepackage{hyperref}       
\usepackage{url}            
\usepackage{authblk}
\usepackage{booktabs}       
\usepackage{amsfonts}       
\usepackage{nicefrac}       
\usepackage{microtype}      
\usepackage{lipsum}
\usepackage{fancyhdr}       
\usepackage{graphicx}       
\usepackage{indentfirst}
\graphicspath{{media/}}     

\title{Implementing Digital Twin in Field-Deployed Optical Networks: Uncertain Factors, Operational Guidance, and Field-Trial Demonstration
}

\author[a]{Yuchen Song}
\author[a]{Min Zhang}
\author[a]{Yao Zhang}
\author[b]{Yan Shi}
\author[b]{Shikui Shen}
\author[a]{Bingli Guo}
\author[a]{Shanguo Huang}
\author[a]{Danshi Wang*}
\affil[a]{State Key Laboratory of Information Photonics and Optical Communications, Beijing University of Posts and Telecommunications, Beijing 100876, China}
\affil[b]{China Unicom Research Institute, Beijing, China}
\affil[*]{Email:danshi\_wang@bupt.edu.cn}

\begin{document}
\maketitle

\begin{abstract}
Digital twin has revolutionized optical communication networks by enabling their full life-cycle management, including design, troubleshooting, optimization, upgrade, and prediction. While extensive literature exists on frameworks, standards, and applications of digital twin, there is a pressing need in implementing digital twin in field-deployed optical networks operating in real-world environments, as opposed to controlled laboratory settings. This paper addresses this challenge by examining the uncertain factors behind the inaccuracy of digital twin in field-deployed optical networks from three main challenges and proposing operational guidance for implementing accurate digital twin in field-deployed optical networks. Through the proposed guidance, we demonstrate the effective implementation of digital twin in a field-trial C+L-band optical transmission link, showcasing its capabilities in performance recovery in a fiber cut scenario.
\end{abstract}


\section{Introduction}
\noindent Communication networks supports the exponential surge in data traffic, driven by factors ranging from the proliferation of social media platforms to the escalating demand of high-performance computing. As networks continue to expand in size and complexity, their behavior becomes increasingly intricate, posing challenges in their management. The concept of digital twin (DT) has been introduced into communication networks with the aim of creating the virtual replicas of physical network entities in cyber space to facilitate their high-reliability operation and efficient management, ultimately driving the networks towards automation \cite{RN624, RN621}.

Among communication networks, optical networks serve as the fundamental infrastructure for supporting high-speed, large-capacity, and long-haul data transmission. They face distinctive challenges due to their wide geographical coverage, complex traffic routes management, and the presence of significant optical nonlinearities especially in wideband scenarios \cite{RN275}. In response to these challenges, numerous studies have focused on developing frameworks, standards, and applications for the Digital twin of optical network (DTON), demonstrating its effectiveness through simulations and laboratory experiments under near-ideal conditions \cite{RN538, RN531}. However, to the best of our knowledge, few studies have been reported on the effective implementation of DTON in field-deployed networks, which is crucial for practical applications of DTON to achieve tangible benefits in terms of capital expenditure (CAPEX). 

Implementing accurate DTON in field-deployed networks presents distinct challenges when compared to well-conditioned simulations and laboratory experiments in multiple practical issues as summarized in Fig. 1. These issues start from imperfect environments, encompassing factors such as fluctuating temperatures and pressures, which introduce disturbances to device behaviors by affecting their physical mechanisms and parameters. Another issue arises from the diverse vendor-specific control functions present in different commercial devices, complicating the modeling of these devices. Furthermore, unexpected human activities and the ageing of network devices can lead to deviations in network status from recorded data, without prompt updates. The extensive geographical coverage of field-deployed networks, coupled with the need to maintain uninterrupted data transmission, makes it challenge to conduct detailed measurements on multiple devices. Additionally, the inherent manufacturing variability among devices means that each device is different. This aspect becomes especially problematic in multi-vendor operations, where achieving accurate characterization of all deployed devices becomes exceptionally challenging. These issues outlined in Fig. 1 present obstacles to the successful implementation of DTON in field-deployed networks. However, a comprehensive and in-depth discussion of these issues is currently lacking in the literature. Current studies on the implementation of DTON in field-deployed networks either overlook these imperfect conditions \cite{RN577} or predominantly focus on resolving particular issues \cite{RN293}. There is a pressing need for practical operational guidance to address these complex challenges effectively.
\begin{figure}
  \centering
  \includegraphics[width=1\linewidth]{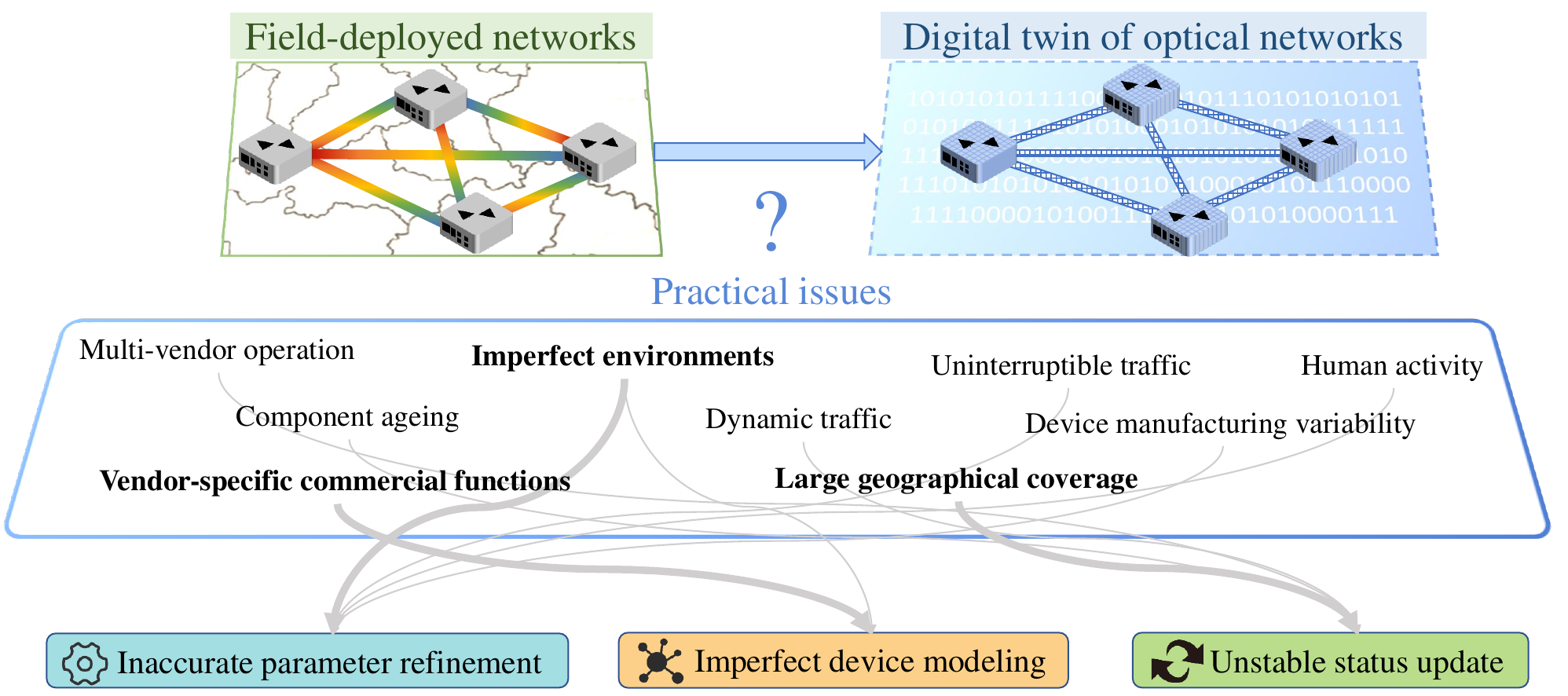}
  \caption{Practical issues with connections to three main challenges of field-deployed optical networks.}
  \label{F1}
\end{figure}

In this article, we conduct a comprehensive discussion on why are DTONs prone to be inaccurate in field-deployed optical networks from various practical issues and conclude them into three primary challenges: inaccurate parameter refinement, imperfect device modeling, and unstable status updating. We evaluate the importance of factors inside these challenges from two indexes. Building upon these analyses, an operational guidance with potential solutions on how to implement DTON in field-deployed optical networks is proposed. Finally, we demonstrate the proposed guidance on the DTON implementation of a field-trial wideband wavelength-division multiplexing (WDM) system operating on C+L-band. A fiber cut scenario is set up, and the performance optimization and recovery based on the established accurate DTON is verified. 
The main contributions of this work can be summarized as below:
\begin{itemize}
    \item Discussion of the uncertain factors on why is DTON prone to be inaccurate in field-deployed networks.
    \item An operational guidance with potential solutions on how to implement DTON in field-deployed networks.
    \item The operational guidance on implementing DTON is validated on a field-trial C+L-band transmission link.
    \item Performance recovery enabled by the implemented DTON is tested on a fiber cut scenario.
    \item A physics-informed machine learning method is used for the simultaneous identification of critical parameters.
\end{itemize}

\section{UNCERTAIN FACTORS – WHY IS DIGITAL TWIN PRONE TO BE INACCURATE IN FIELD-DEPLOYED NETWORKS}
\noindent These practical issues can have diverse impacts in field-deployed networks, necessitating a systematic categorization for efficient solutions. There are three main elements in networks: devices, parameters, and network status. For various devices, characterizing their behaviors demands distinct knowledge across different physical domains. Accurate device modeling is a fundamental requirement, and precise parameter identification is crucial for achieving accurate network outputs. These parameters include device characteristics related to fiber types, amplifier gain, noise figures, and more, which typically cannot be manually adjusted post-manufacturing. The status of deployed optical networks includes network topology, route and spectrum assignment, channel power and performance, and device configuration. Network status is expected to be manually adjusted during regular operation. To better understand and address the intricate web of practical issues, we have categorized their impacts into three main problems: inaccurate parameter identification, imperfect device modeling, and unstable status updating. In Fig. 1, we have visually represented these connections, with the width of the lines indicating the strength of their impact on the corresponding challenge. 

To conduct a qualitative assessment of the impact of different factors within these three problems, we propose two metrics: the DT-index and the Field-index, as depicted in Fig. 2. The DT-index quantifies the significance of a factor in terms of its impact on the accuracy of the DTON. A higher DT-index indicates that a factor has a more pronounced effect on the precision and reliability of the DTON. In other words, it measures how crucial a factor is for ensuring the accuracy of the digital twin representation. A higher Field-index suggests that a factor introduces greater uncertainty or dynamicity when applied in real-world network scenarios. Factors with a higher Field-index demand heightened attention and consideration when implementing DTON in field-deployed networks. These two indexes offer valuable guidance for decision-makers involved in the implementation of DTON. These indexes assist in making informed trade-offs and focusing efforts on the most critical factors, thereby enhancing the effectiveness of DTON in real-world applications.
.

\begin{figure}
  \centering
  \includegraphics[width=1\linewidth]{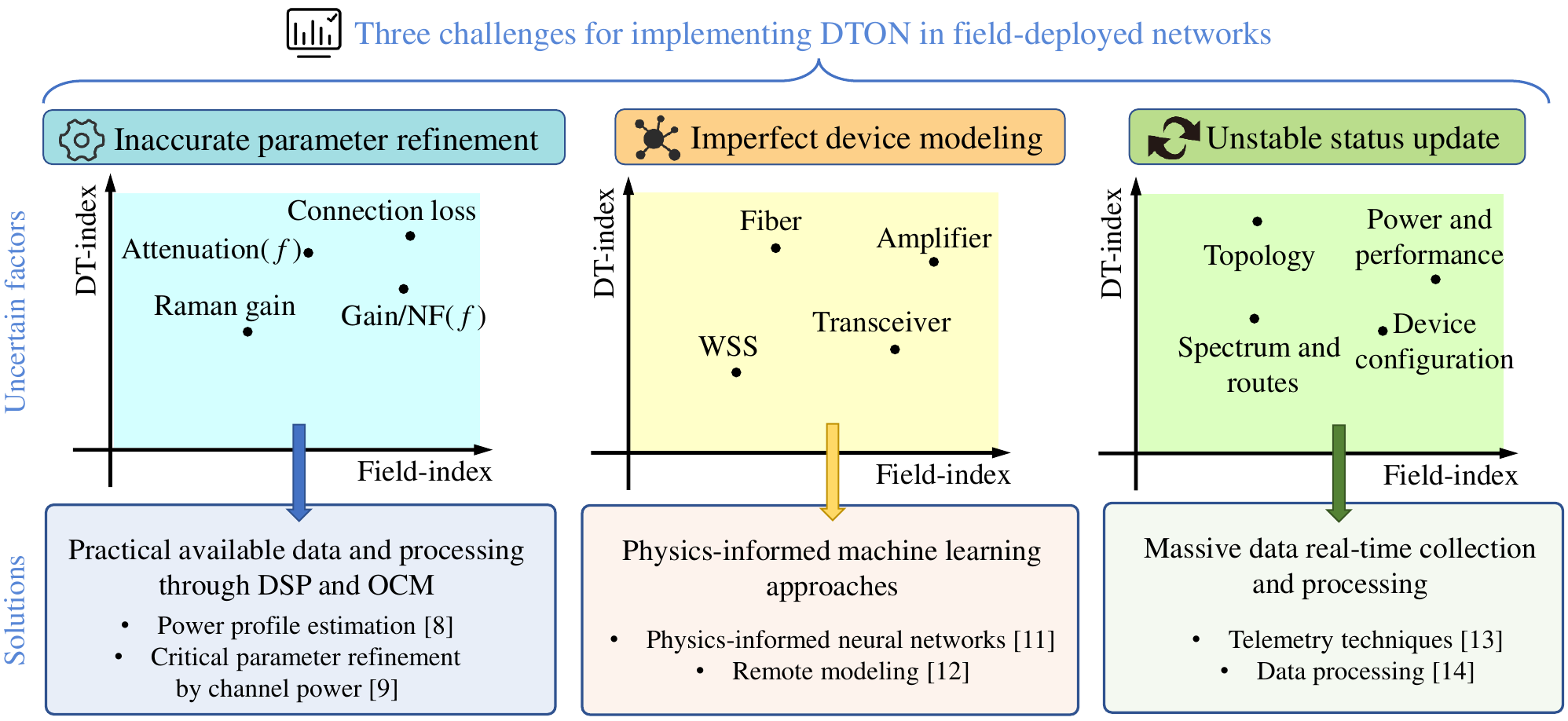}
  \caption{Three main challenges with corresponding uncertain factors and potential solutions. These uncertain factors are evaluated by the DT- and Field-index.}
  \label{F2}
\end{figure}

\subsection{Inaccurate system parameter}
\noindent While some basic parameters can be obtained from the manufacturer's documentation or deployment datasheets, field-deployed networks introduce a level of uncertainty due to imperfect environments, human activities, and device ageing. These factors can contribute to additional losses and noises within the network, making certain system parameters less reliable. Furthermore, some critical parameters, such as the connector loss between the fiber and amplifier, may not be readily available in datasheets, as depicted in Fig. 2. The connector loss occurring before the fiber can lead to a reduction in input fiber power, significantly impacting the accumulated nonlinearity. Similarly, the connector loss preceding the erbium-doped fiber amplifiers (EDFA) can affect both the gain and the Amplified spontaneous emission (ASE) noise level. Thus, factor of connector loss exhibits both large DT and Field index and should be carefully identified in implementing DTON.

As optical networks evolve from the conventional C band to C+L band, the accuracy of frequency-dependent parameters becomes increasingly important. Examples of such parameters include the attenuation spectrum, Raman gain spectrum, and amplifier gain profile. Of these factors, the amplifier gain and noise figure (NF) profile is particularly uncertain in field networks and can exert a significant influence on the ultimate accuracy of the DT. On the other hand, the Raman gain spectrum tends to be relatively stable and primarily impacts wideband transmission scenarios.

Unlike laboratory environments where comprehensive measurements can be readily conducted, field-deployed networks cover vast geographical areas and carry live user traffic. Therefore, techniques for parameter identification and refinement must be conducted remotely without causing any disruption to ongoing network operations. In this context, longitudinal parameter identification using digital signal processing (DSP) techniques proves to be highly valuable in detecting anomalous losses \cite{RN552}. Critical parameters like connector loss, Raman gain strength, and gain profile can be refined with the assistance of an optical channel monitor (OCM) \cite{ecoc}. This practical approach is well-suited for application in field-deployed networks, allowing for real-time refinement without interrupting network operations.

\subsection{Imperfect device modeling}
\noindent The imperfect device modeling means the device behavior cannot be fully captured and well characterized by an explicit model. Imperfect environmental conditions and various vendor-specific functions in different commercial devices can significantly influence the behavior of deployed devices, posing challenges on their modeling. A prime example of this is the EDFA, which poses modeling difficulties due to its intricate input-dependent mechanism and a wide array of vendor-specific functions in commercial devices. The EDFA factor exhibits a high Field-index as shown in Fig. 2 due to its dynamic nature and the lack of a comprehensive model and a high DT-index due to its large impact on both linear and nonlinear impairments along the link. Another factor contributing to DT inaccuracy is the filtering effect induced by cross-connections, such as the wavelength selective switch (WSS), in optical networks. The impact of these two factors will accumulate with the increasement of transmitting spans. For field-deployed transponder, additional linear and nonlinear impairments can be introduced and can be the main impairment source in transmissions with short distance.

On the other hand, the fiber channel, which is the primary source of impairments in optical networks, can be well characterized using various methods. These include solving the nonlinear Schrödinger equation (NLSE) using split-step methods and utilizing the Gaussian noise (GN) model for quality of transmission (QoT) estimation \cite{RN133}. Both methods have undergone extensive validation, resulting in a relatively low field index as shown in Fig. 2. When it comes to field-deployed devices, remote modeling becomes essential since physically accessing each device and collecting extensive data is impractical. Due to the inherent uncertainties, modeling through data-driven neural networks (NNs) has proven to be an effective approach and has been extensively validated, especially in the case of EDFA modeling \cite{RN542}. However, it's important to note that data-driven modeling can be somewhat opaque and challenging to interpret. In such cases, a hybrid approach that combines physical principles with data-driven modeling holds promise as a technique for device modeling in field-deployed networks, such as physics-informed machine learning \cite{RN26}.

\subsection{Unstable status update}
\noindent Real-time calibration of DTON with its physical counterpart becomes crucial for fulfilling its role in full-life cycle management. However, status update in deployed networks can be unstable, including delayed or even lost status, due to device ageing and unexpected human activities. The update of route and spectrum assignment (RSA), whether it is caused by manual adjustment or unforeseen fiber cuts, is a regular operation happening frequently. These update of RSA with a large DT index have a significant impact on transmission performance and provide vital information for network configuration management, such as amplifier configuration and dummy channel filling. It should be noted that in wideband systems, the dynamicity due to RSA update is more profound due to frequency-dependent effects such as stimulated Raman scattering (SRS) effect. The update of device configuration and network topology can generally be executed relatively quickly in field-deployed networks, leading to a relatively lower DT- and Field-index. The per-channel power and performance, which are essential for calibrating the DTON, carry a high DT index. Nevertheless, maintaining the stability of these updates can be problematic, especially when unexpected events occur.

The stable status update in field-deployed networks requires the development of telemetry techniques \cite{RN548}. Furthermore, the utilization of DSP and machine learning techniques opens up possibilities for gathering more information \cite{RN550}. Additionally, the adoption of open and disaggregated management practices is essential for effectively handling multi-vendor device operations.

\section{OPERATIONAL GUIDANCE – HOW TO IMPLEMENT ACCURATE DIGITAL TWIN IN FIELD-DEPLOYED OPTICAL NETWORKS }
\noindent Based on the aforementioned analysis, we aim to provide an operational guidance on the implementation workflow of DTONs with focus on these factors with high DT and Field index. While it may not offer an exhaustive set of standard steps, this operational guidance aims to provide some key guidelines in the implementation process.

As illustrated in Fig. 3, at the first step, data of network status including topology, device configuration, spectrum occupation and routes are collected. Additionally, basic device parameters can be obtained from the manufacturer's documentation, such as fiber and amplifier characterization. It should be noted the back-to-back bit error rate (BER) versus signal-to-noise ratio (SNR) for transponder is necessary for translating the traffic bit-error-ratio (BER) to the final generalized signal-to-noise ratio (GSNR). The GSNR can be derived from GN models, accounting for both linear and nonlinear impairments. Basic device model without frequency or input-dependent characteristics are applied. 

\begin{figure}
  \centering
  \includegraphics[width=1\linewidth]{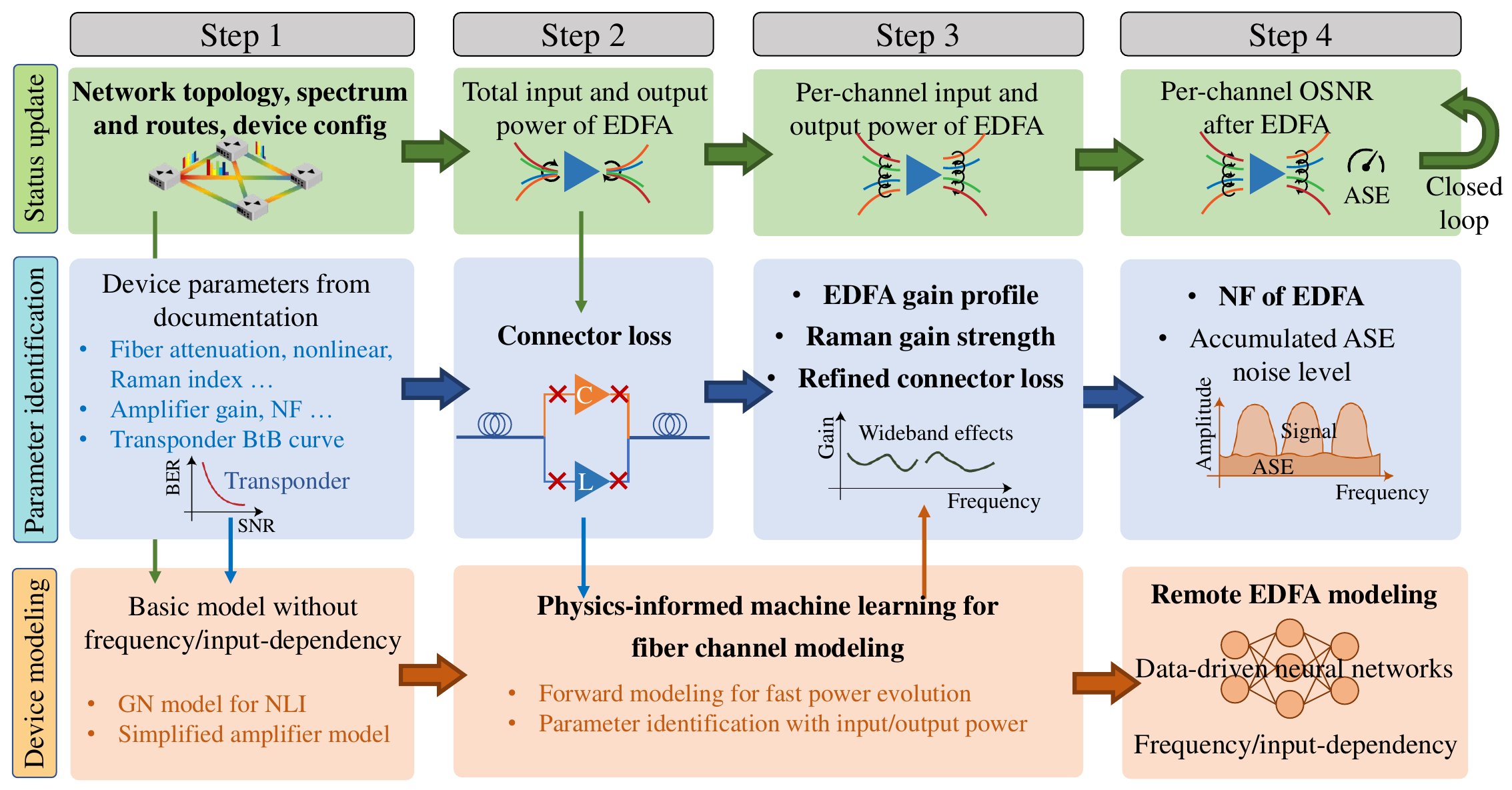}
  \caption{Operational guidance on implementing DTON in field-deployed networks. Factors with high DT- and Field-index in boldface.}
  \label{F3}
\end{figure}

Traditionally, in well-controlled laboratory settings, this first step dataset, coupled with basic device models, is often sufficient to create an accurate DTON. This is the only step most commonly used in current field-deployed networks [6], which we denote as DTON\#1. However, as indicated by the results in the Fig. 5, this approach falls short of ensuring accuracy. Consequently, additional steps are required to enhance the precision of the DTON in practical field-deployed networks. 

In step 2, we delve into querying essential network status information – the total power at the input and output of the EDFA. The primary objective of this step is to infer the connection losses, which is a critical factor with high DT- and Field-index. At this step, equal loss can be assumed for the input and output of fiber.	Several prior studies focused on the implementation of DTON in field-deployed networks have noticed issues related to inaccurate parameters. These studies have attempted to improve the DTON accuracy by coarsely refining connector loss \cite{RN293}. This situation closely resembles the approach outlined in our proposed step 2 of the operational guidance, which we denote as DTON\#2.

Moving to step 3, the network status of per-channel power is updated, which aims at abstracting frequency-dependent effects. With this information, the static gain profile of EDFA and the strength of SRS effects can be estimated. Moreover, among step 2 and 3, the implementation of physics-informed neural networks (PINNs) can be particularly valuable for modeling the fiber channel. PINNs utilize the NLSE as a regularization, ensuring that the modeling is grounded in physical principles. The simulation of multi-channel power evolution can be speeded up with physical guarantee. With the data of per-channel input/output power of fiber, PINNs can refine the connector loss before and after fiber with higher accuracy, while also refine the wideband attenuation spectrum and SRS strength \cite{ecoc}, as illustrated in \ref{F3}. 
\begin{figure}
  \centering
  \includegraphics[width=0.8\linewidth]{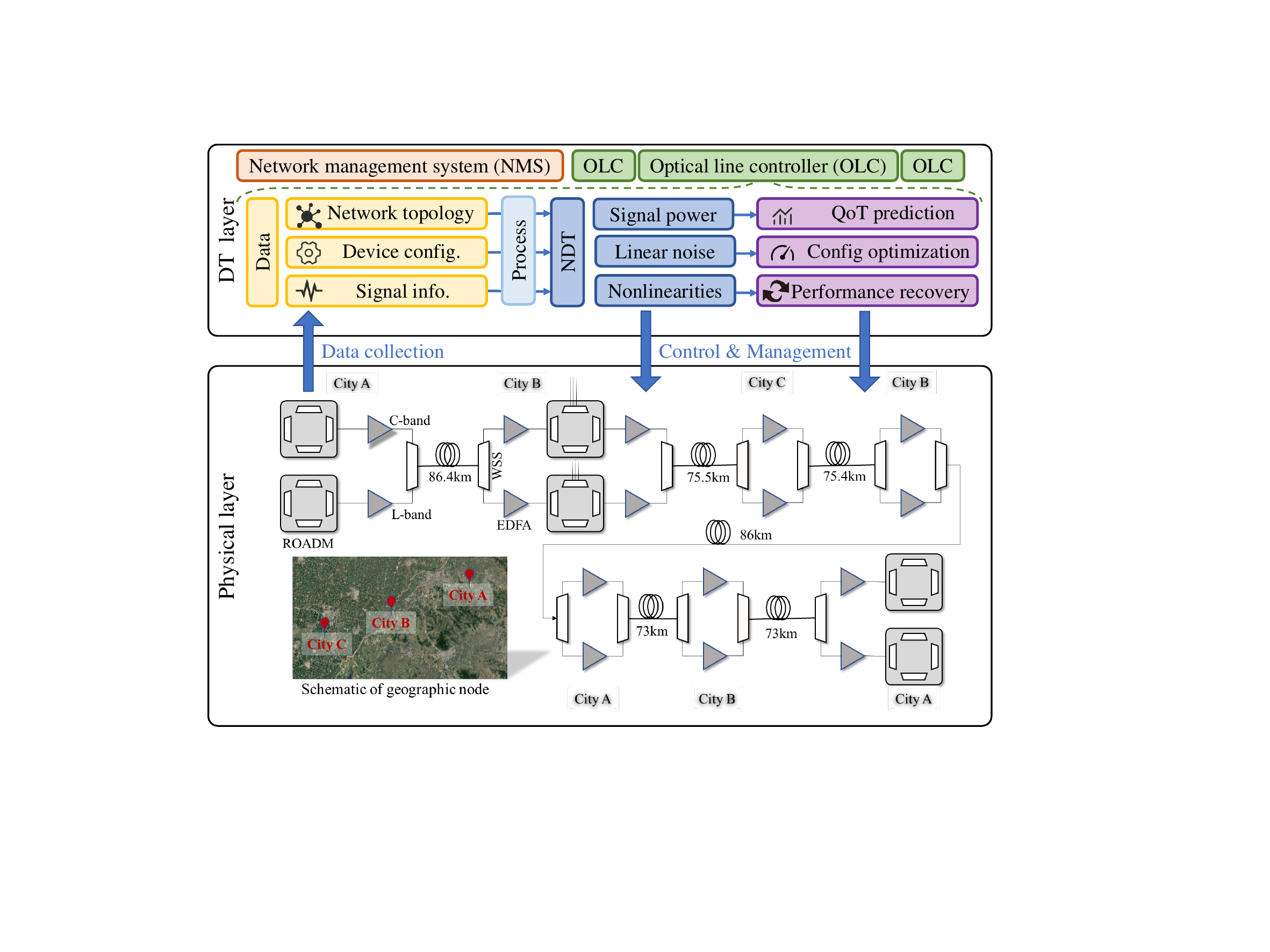}
  \caption{Setup of field-trial C+L-band link and the schematic of the implemented DTON.}
  \label{F4}
\end{figure}

The fourth step requires the information of optical SNR (OSNR) to measure the ASE noise level of each EDFA, which is necessary for accurately estimating the final performance as the linear noise dominates around the optimal working point of optical networks \cite{RN133}. With the information of OSNR before and after each EDFA, the NF can be calculated. A comprehensive EDFA modeling considering the complex input-dependent behavior and gain-NF relationship may require big data collection. It is possible to conduct modeling for EDFA using data-driven NNs in the remote way \cite{RN548}. 
However, there is a chance that some deployed EDFA is not equipped with optical channel monitor (OCM), which is necessary for monitoring per channel power and OSNR. In this case, a compromise method is to just monitor the final OSNR at receiver side. As the linear noise is the big part of the final impairments, this step is necessary for the performance estimation, as shown in Fig. 3. The final step for verification is to change the loading conditions to check if the nonlinearity is estimated correctly. If not, it is highly possible that some connection losses are inaccurate, which requires further refinement and calibration step by step . 

When the DTON is implemented accurately, closed-loop network status update is required to ensure the synchronous update with physical networks. To overcome the problem of unstable status updating, various data collection techniques, such as telemetry, play a crucial role. The advancements in coherent digital signal processing (DSP) and machine learning techniques have enabled feasible data collection and monitoring capabilities, enhancing the accuracy and real-time nature of DTON.

\section{FIELD-TRIAL DEMONSTRATION ON C+L-BAND TRANSMISSION LINK}
\noindent A field-trial transmission link in China Unicom’s metro optical transport networks was selected to validate the efficacy of the operational guidance on implementing DTON, as depicted in Fig. 4. The status information can be queried from network management system (NMS).
\subsection{Field-trial C+L-band links}
\noindent The field-trial C48+L48 WDM transmission link consists of six amplified spans with a maximum length of 86.4km (totaling 469.3km of G.652 SMF). A ROADM station is placed at the second site in city B, while in other sites, in-line EDFA is used for separated amplification of C- and L-band. OCM is placed at the front and end of EDFA for channel power collection. Three commercial 400Gb/s transponders on the C-band and two on the L-band are configured for five channels under test (CUT), and probabilistic constellation shaping-quadrature amplitude modulation (PCS-16QAM) with 91.6 baud rate is modulated for optical transmission with 100GHz channel spacing. In the transceiver side, signals are Mux/Demux by ROADM, and other channels are filled with filtered ASE noise for full load configuration on C+L-band. The transmission bandwidth occupies the L-band, from 186.1 THz to 190.8 THz, and the C-band, from 191.4 THz to 196.1 THz with a total of 96 channels. It should be noted that although some nodes are located at the same city, they are not in the same station. 
\subsection{Implementation of DTON for field-trial link and performance comparison}
\noindent The implemented DTON contains the signal power propagation model considering the SRS, the EDFA model with the linear noise, and the GN model for the estimation of nonlinearities, as illustrated in Fig. 4. At the first step, the DTON can be implemented with the information of basic network status, device parameter, and device model. It can be observed in Fig. 5 that the results of step 1 denoted by blue square deviate from the target field measurements. It should be noted that the step 1 is also the most common used approach of DTON implementation, which will lead to the implementation of DTON\#1 as defined in section. III. 

\begin{figure}
  \centering
  \includegraphics[width=0.4\linewidth]{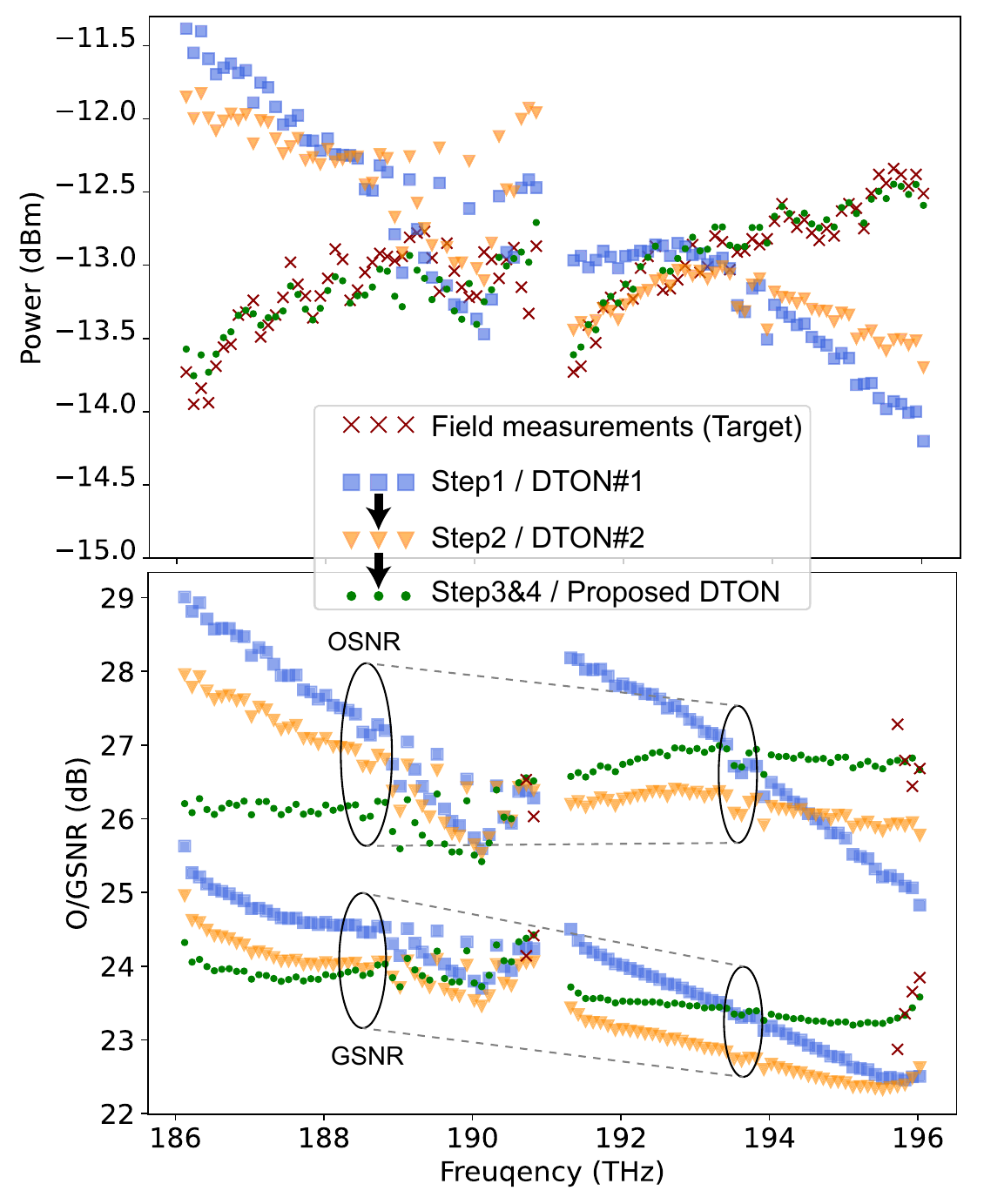}
  \caption{Results of implemented DTON on the prediction of channel power and O/GSNR at different steps. These results also show comparison of the proposed DTON to DTON\#1 and \#2.}
  \label{F5}
\end{figure}

Following the steps illustrated in Fig. 3, we first queried the total input/output power of deployed EDFA from inside power module or OCM. In each span, connector loss was estimated considering the number of connections and the difference between total loss and fiber attenuation. It should be noted that for field-deployed optical networks, there may be a large number of connections, which can result in large connection loss at a point. After step 2, the uncertainty of connection loss is reduced, and thus the estimation of nonlinearity and particularly the SRS is more accurate by DTON. From the results of step 2 shown in Fig. 5, it can be seen that both channel power and O/GSNR results came closer to the target field measurements. The methodology in steps 1 and 2 closely resembles the scenario in which connector loss is roughly estimated during the implementation of DTON in field-deployed networks, previously referred to as DTON\#2.

For step 3, we queried the information of channel power at each node from OCM. The static EDFA gain profiles were abstracted from the difference of channel power before and after EDFA. In this step, the connector loss, gain profile, and SRS strength can be identified using the PINNs [9]. On the following step, we queried the OSNR after each EDFA and estimated the NF, which determines the level of ASE noise. 

Finally, it can be observed in the results of step 3\&4 in Fig. 5 that with refined EDFA gain profile and more accurate NF, both the channel powers and O/GSNR estimated by the improved DTON match well with the target field measured ones. Furthermore, the results of step 3\&4 – our proposed DTON, exhibit a higher level of accuracy compared to DTON\#1 and \#2, with an improvement of 2.3 and 1.8 dB for channel power at its maximum. In the case of GSNR, the proposed DTON surpasses DTON\#1 and \#2 by 1.3 and 0.6 dB, respectively. These results underscore the effectiveness of our proposed operational guidance.
\begin{figure}
  \centering
  \includegraphics[width=0.4\linewidth]{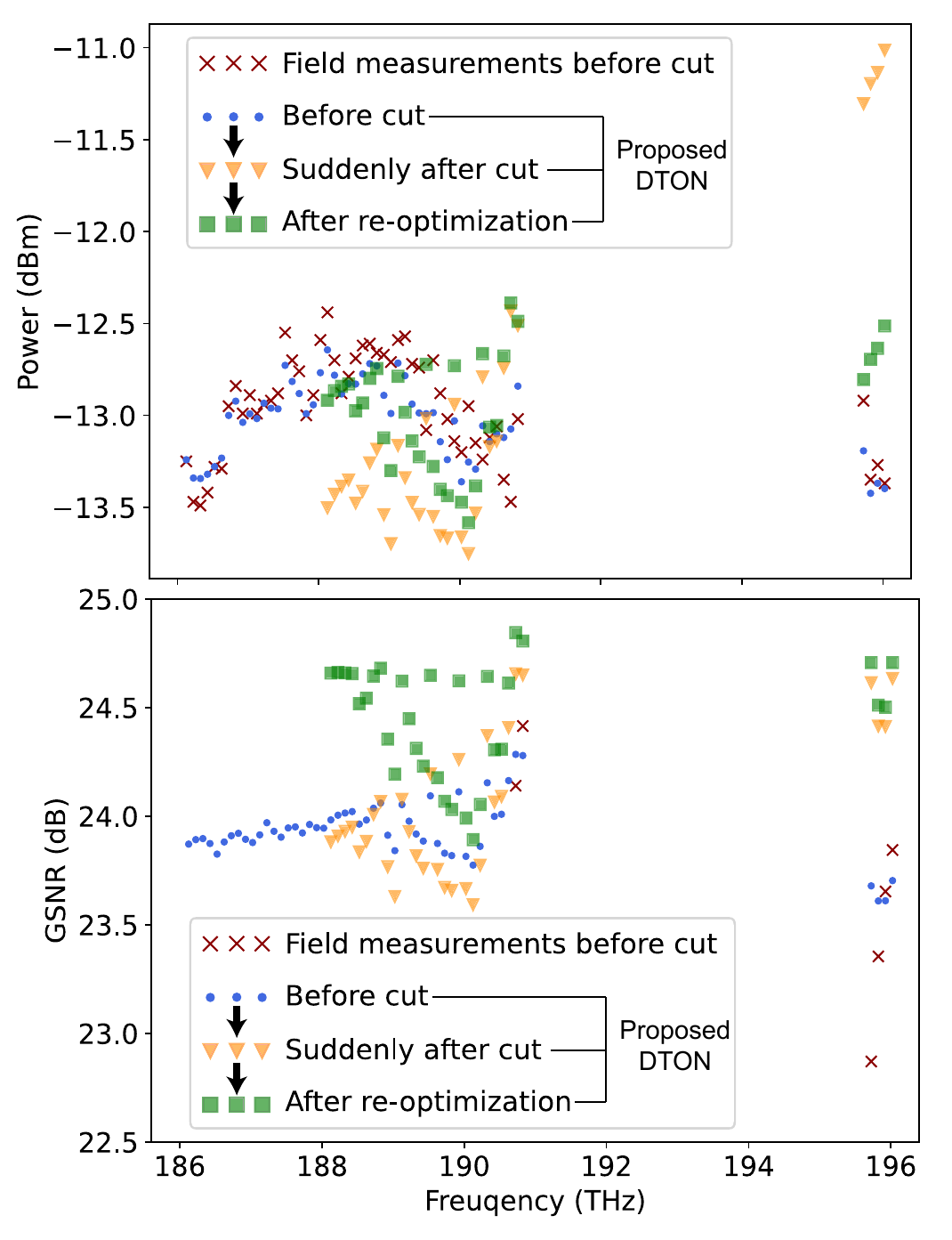}
  \caption{Performance recovery for channel power and GSNR of implemented DTON on a fiber cut scenario.}
  \label{F6}
\end{figure}
\subsection{Optimization using DTON for field-trial link}
\noindent Based on the accurate implemented DTON, we explore its practical application in field-deployed optical networks for optimizing and recovering link performance by adjusting EDFA configurations (i.e., gain and tilt). In order to further increase the reliability of DTON, we queried per channel powers at the input and output of each EDFA under different configurations. This enables us to establish a simple EDFA model considering the dependence between gain and NF interpolate different gain profiles for varying gain tilts [15]. We select the multi-step gradient-ascent optimization algorithm validated by experiments in \cite{RN25}. Before the operation of DTON, its accuracy should be validated under different loadings and EDFA configurations of the link.

As no important user traffic is carried on this field-trial link, it is possible to manipulate the loadings. Thus, we set up a scenario, where the link is initially filled with full L-band channels and three CUT on the C-band with optimized EDFA configurations, and then a sudden fiber cut on the previous section of this network causes the dropping of 20 channel at the low frequency end of L-band. As depicted in Fig. 6, the sudden drop of these 20 channels at the high frequency reduces the strength of SRS, resulting in a more positive tilt in the channel power after the cut. This effect occurs because the EDFA configuration is optimized for the full L-band loading to counteract stronger SRS. The imbalanced channel power along the link alters the accumulated impairments, including linear impairments from EDFA and nonlinearities from the fiber, across the transmission bandwidth. As a result, an imbalance in the O/GSNR is observed.

Upon detecting this abnormal performance reduction, DTON swiftly conduct EDFA configuration re-optimization to restore the balanced performance. After re-optimization, the channel powers become much more balanced, and the GSNR improves due to fewer channels and consequently reduced nonlinearity.

\section{Conclusion}
\noindent In this article, we focused on the implementation of accurate DT in field-deployed optical networks. The reason why the DTON prone to inaccuracy was explained from the aspect of uncertain factors of imperfect component modeling, delayed status updating, and inaccurate system parameter. The consequence of inaccurate DTON resulted from different uncertain factors was analyzed. The operational guidance for implementing DT in field-deployed optical networks was proposed with focus on solving the problem of inaccurate system parameters. The effectiveness of our proposed guidance was demonstrated through its successful application on a field-trial C+L-band WDM link, resulting in a substantial improvement in DTON accuracy. Additionally, we set up a fiber cut scenario and showcased the EDFA configuration optimization based on established accurate DTON. The comprehensive analysis of uncertain factors and the proposed operational guidance in this article offer practical insights and solutions on the application of DT to field-deployed optical networks and bring DTON closer to reality.
To implement DTON in field-deployed networks and realize its benefits throughout their lifecycle, additional efforts are essential to address these three key challenges. The modeling of devices needs to become more efficient to capture a broader spectrum of information during signal propagation. Currently, the detailed calculation of wideband signal propagation in fiber remains a complex endeavor, as does the accurate modeling of EDFA under dynamic loading conditions. For the refinement of numerous critical parameters, including anomaly lump loss, amplifier gain profiles, filtering impairments, and more, the development of practical methods with reduced complexity is of paramount importance. Furthermore, telemetry techniques are advancing, enabling the collection of more data. However, the efficient management and processing of this data within open and disaggregated optical networks remain an open problem. In conclusion, optical networks are evolving towards increased intelligence, and the implementation of DTON plays a pivotal role in this transformation.

\section*{Acknowledgments}
This was was supported in part by National Natural Science Foundation of China No. 62171053, Beijing Nova Program No. 20230484331, and BUPT Excellent Ph.D. Students Foundation No. CX2022123.

\bibliographystyle{unsrt}  
\bibliography{references}

\end{document}